\documentclass[traditabstract]{aa} 
\usepackage{graphicx}
\usepackage{txfonts}
\usepackage{fontawesome}
\usepackage{mathtools}
\usepackage{microtype}
\usepackage{graphicx}
\usepackage{subfigure}
\usepackage{caption}
\usepackage{subcaption}
\usepackage{booktabs} 
\usepackage{subcaption}
\usepackage{fontawesome}
\usepackage{hyperref}
\usepackage{xcolor}
\usepackage{multirow}
\usepackage{bigdelim}
\usepackage{float}
\usepackage{algorithm}
\usepackage{algorithmic}

\begin{document} 

   \title{Joint multiband deconvolution for Euclid and Vera C. Rubin images}

   \author{U. Akhaury
          \inst{1}
          \and 
          P. Jablonka
          \inst{1,2}
          \and
          F. Courbin
          \inst{1,3,4}
          \and
          J.-L. Starck
          \inst{5,6}
          }
   \institute{Laboratory of Astrophysics, Ecole Polytechnique Fédérale de Lausanne (EPFL), Observatoire de Sauverny, CH-1290 Versoix, Switzerland.
              \email{utsav.akhaury@epfl.ch}
             \and
             GEPI, Observatoire de Paris, Université PSL, CNRS, 5 Place Jules Janssen, 92190 Meudon, France.
             \and
             ICC-UB Institut de Ci\`encies del Cosmos, Universitat de Barcelona, Mart\'i Franqu\`es, 1, E-08028 Barcelona, Spain.
             \and
             ICREA, Pg. Llu\'is Companys 23, Barcelona, E-08010, Spain.
             \and
             Université Paris-Saclay, Université Paris Cité, CEA, CNRS, AIM, 91191, Gif-sur-Yvette, France
             \and
             Institutes of Computer Science and Astrophysics, Foundation for Research and Technology Hellas (FORTH), Greece \\
             }

   \date{Received September 15, 1996; accepted March 16, 1997}

  \abstract
    {With the advent of surveys like Euclid and Vera C. Rubin, astrophysicists will have access to both deep, high-resolution images and multiband images. However, these two types are not simultaneously available in any single dataset. It is therefore vital to devise image deconvolution algorithms that exploit the best of both worlds and that can jointly analyze datasets spanning a range of resolutions and wavelengths. In this work we introduce a novel multiband deconvolution technique aimed at improving the resolution of ground-based astronomical images by leveraging higher-resolution space-based observations. The method capitalizes on the fortunate fact that the Rubin $r$, $i$, and $z$ bands lie within the Euclid VIS band. The algorithm jointly de-convolves all the data to convert the $r$-, $i$-, and $z$-band Rubin images to the resolution of Euclid by leveraging the correlations between the different bands. We also investigate the performance of deep-learning-based denoising with DRUNet to further improve the results. We illustrate the effectiveness of our method in terms of resolution and morphology recovery, flux preservation, and generalization to different noise levels. This approach extends beyond the specific Euclid-Rubin combination, offering a versatile solution to improving the resolution of ground-based images in multiple photometric bands by jointly using any space-based images with overlapping filters.}


   \maketitle

\section{Introduction}

High spatial resolution, a high signal-to-noise ratio (S/N), and broad wavelength coverage are all essential for most observations in astrophysics. However, it is difficult, or even impossible, to have all three simultaneously. Space telescopes, although free of atmospheric turbulence, are limited in size. Ground-based telescopes can deliver high-S/N data but are affected by atmospheric turbulence and have a higher sky background. In addition, blurring by the instrumental or atmospheric point spread function (PSF) differs for each type of data and varies from band to band. To capitalize on the strengths of all types of telescopes and data, it is crucial to develop robust deconvolution techniques that can remove blurring by the PSF and optimize the S/N of the final reconstruction, by combining all observations and accounting for the different bands and PSFs.

Due to the presence of noise, image deconvolution is a challenging ill-posed inverse problem that requires regularization for a well-defined solution. Early approaches in the field acknowledged this issue, proposing solutions such as minimizing the Tikhonov function \citep{tikhonov1977solutions} or maximizing the entropy of the solution \citep{skilling1984}. Bayesian methods also emerged, including the Richardson-Lucy algorithm applied to early \textit{Hubble} Space Telescope (HST) data \citep{Richardson1972, Lucy1974}. A novel approach proposed by \citet{MCS} separated point sources from extended ones and used a narrow PSF for deconvolution to achieve an improved resolution suitable for the chosen pixel sampling. This approach was improved with wavelet regularization for the extended channel \citep{firedec} and further refined by \citet[STARRED;][]{STARRED}, who employed Starlets, an isotropic wavelet basis \citep{Starck2015}, to regularize the solution. There have also been efforts to jointly deconvolve multiple astronomical observations of the same sky region \citep{Donath2023Jolideco}. Furthermore, \cite{rldec} explored the combination of multiple sources by demonstrating the application of Richardson-Lucy deconvolution to merge high-resolution, high-noise images with low-resolution, low-noise images. 

A notable advancement in astronomical deconvolution was the use of deep learning. Once trained, deep-learning-based methods offer significant computational efficiency compared to traditional approaches. U-Nets \citep{ronnenberger2015unet} have gained popularity for their nonlinear processing capabilities and multi-scale architecture. Expanding on U-Nets, \cite{sureau2020} introduced the Tikhonet method for deconvolving optical galaxy images, demonstrating its superior performance over sparse regularization methods in terms of the mean squared error and a shape criterion that assesses galaxy ellipticity. \citet{shapenet} improved Tikhonet by incorporating a shape constraint into the loss function. Another powerful architecture, Learnlet \citep{ramzi:hal-03346892}, combines the strengths of wavelets and U-Nets while offering a fully interpretable neural network with minimal hallucination. In our previous work \citep{akhaury2022, sunet_akhaury}, we proposed a two-step deconvolution framework and investigated the performance of convolutional neural network (CNN) and transformer-based denoisers. We concluded that a Swin-transformer-based U-Net (SUNet; \citealt{sunet}) outperforms a CNN-based U-Net \citep{ronnenberger2015unet} in terms of normalized mean squared error (NMSE) and structural similarity index measure.

While deconvolution is primarily used to reconstruct galaxy images at high spatial resolution in each photometric band independently, there are scenarios, particularly at low S/Ns, where joint multiband deconvolution can enhance the detection and characterization of systems. One such potential application is the joint multiband deconvolution of Euclid and Vera C. Rubin images. The Rubin Observatory is set to deliver a dataset of $500$ petabytes across multiple optical frequency bands, while Euclid will observe images spanning the optical and infrared spectrum. Interestingly, the Euclid VIS band (central frequency $= 715$ nm) overlaps with three of the Rubin filters: $r$, $i$, $z$. As a space-based satellite, Euclid will produce images with sharper details due to its narrower PSF compared to Rubin. A related study by \cite{josephmelchior2021} also involves jointly modeling simulated images that model observations from both Euclid and Rubin.

In this work we present a novel multiband deconvolution technique designed to enhance the resolution of ground-based astronomical images by leveraging higher-resolution space-based observations. Our approach, which focuses on the joint deconvolution of Rubin and Euclid images, effectively exploits the overlapping spectral coverage of the Rubin $r$, $i$, and $z$ bands with the Euclid VIS band. By utilizing the Euclid VIS-band image as a term that provides additional information, our technique ensures that the deconvolved Rubin images retain high spatial resolution and accurate photometric measurements. The integration of deep-learning-based denoising further enhances the quality of the deconvolved outputs, reducing background noise without altering the main structures of the galaxies. We generated realistic Euclid and Rubin simulations from HST cutouts of varying magnitudes extracted from the GOODS-N and GOODS-S surveys \citep{goods}. The simulated Euclid-like VIS-band PSF was obtained from \cite{tobias2022}, and the simulated Rubin-like $r$-, $i$-, and $z$-band PSFs from \cite{lsst_2021}. Our method is effective in terms of resolution recovery, flux preservation, and generalization across different noise levels. Through our joint deconvolution approach, we achieve resolution recovery in simulated Rubin  images close to that of HST, a feat nearly impossible with independent deconvolutions of each photometric band. The potential applications of our method go beyond the Euclid-Rubin pair, providing a flexible solution to enhancing the resolution of ground-based images in multiple photometric bands with any overlapping space-based filter band. This versatility is especially important as large-scale astronomical surveys gather increasing amounts of data, creating a need for effective and reliable deconvolution techniques.

In Sect. \ref{dec_problem} we describe the deconvolution problem and introduce our proposed solution. The methodology for generating our dataset is detailed in Sect. \ref{sec:data_exp}. We then present the outcomes of our deconvolution algorithm in Sect. \ref{results}. Finally, Sect. \ref{concl} presents our conclusions. To support reproducible research, the codes utilized in this article are publicly accessible.

\section{The deconvolution problem}
\label{dec_problem}

\subsection{The forward model}
\label{forward_model1}

For the three Rubin filters, let $\mathbf{y}_r, \mathbf{y}_i$, and $\mathbf{y}_z \in\mathbb{R}^{n\times n}$ be the corresponding observed images and $\mathbf{h}_r, \mathbf{h}_i$, and $\mathbf{h}_z \in\mathbb{R}^{n\times n}$ be the PSFs. If $\mathbf{x}_r^t, \mathbf{x}_i^t$, and $\mathbf{x}_z^t \in \mathbb{R}^{n \times n}$ denote the corresponding target images, $\ast$ denotes the convolution operation, and $\eta_r, \eta_i$, and $ \eta_z \in\mathbb{R}^{n\times n}$ denote additive noise, the observed Rubin images can then be modeled as
\begin{align} 
    \mathbf{y}_r &= \mathbf{h}_r \ast \mathbf{x}_r^t + \eta_r \\[5pt]
    \mathbf{y}_i &= \mathbf{h}_i \ast \mathbf{x}_i^t + \eta_i \\[5pt]
    \mathbf{y}_z &= \mathbf{h}_z \ast \mathbf{x}_z^t + \eta_z. 
\end{align}

As for Euclid, let $\mathbf{y}_{euc} \in\mathbb{R}^{n\times n}$ be the observed image, $\mathbf{x}^t_{euc} \in\mathbb{R}^{n\times n}$ be the target image, and $\mathbf{h}_{euc} \in\mathbb{R}^{n\times n}$ be the PSF. If $\eta_{euc} \in\mathbb{R}^{n\times n}$ denotes additive noise and $\alpha_r, \alpha_i, \alpha_z \in \mathbb{R}^{n}$ denote the corresponding fractional flux contribution from each Rubin filter, the target and the observed images can be modeled as
\begin{align} 
    \mathbf{x}^t_{euc} &= \alpha_r \mathbf{x}^t_r + \alpha_i \mathbf{x}^t_i + \alpha_z \mathbf{x}^t_z \\[5pt]
    \mathbf{y}_{euc} &= \mathbf{h}_{euc} \ast \mathbf{x}_{euc}^t + \eta_{euc.}
\end{align}

\begin{figure}[h!]
    \centering\label{subfig:LSST_filters}\includegraphics[width=\linewidth]{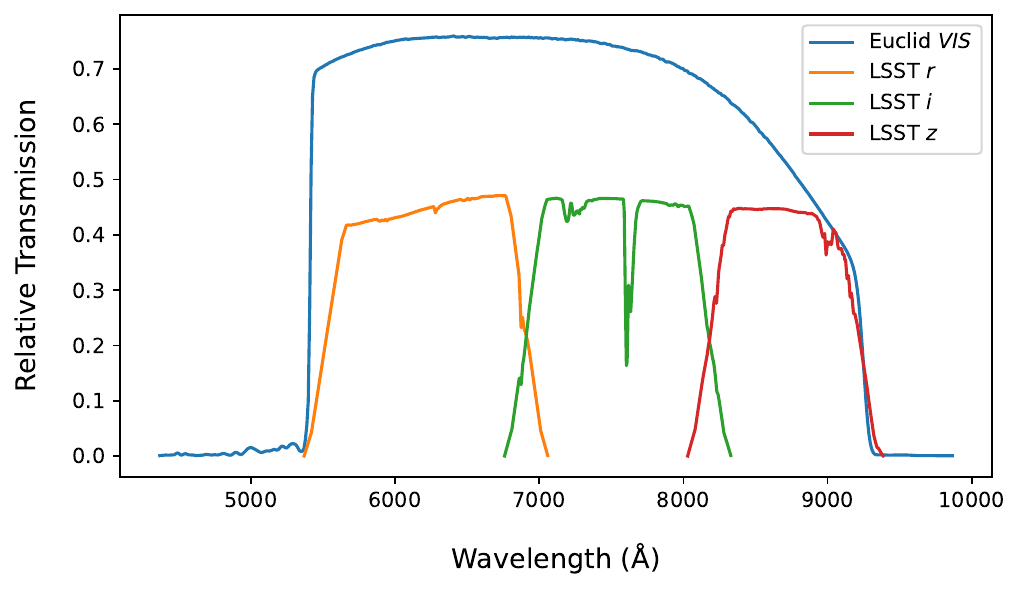}
    \caption{Filter curves for Euclid and Rubin.\  The relative filter transmission is shown as a function of the wavelength. The Euclid VIS band overlaps with three of the Rubin filters: $r$, $i$, and $z$.}
    \label{fig:filters}
\end{figure}

\noindent The motivation behind taking the weighted sum of the Rubin images to model the Euclid image can be seen in Fig. \ref{fig:filters}, which shows the overlap between the Rubin and Euclid filters. 

\subsection{The proposed solution}
\label{forward_model}

We formulated the following loss functions and minimize them using gradient descent to recover the optimal solutions:

\begin{align} 
    L_r(\mathbf{x}_r) &= \frac{1}{2} \left\Vert \frac{\mathbf{h}_r \ast \mathbf{x}_r - \mathbf{y}_r}{\sigma_r} \right\Vert_F^2  + \lambda_{r} \textbf{K} \label{vloss}\\[10pt]
    L_i(\mathbf{x}_i) &= \frac{1}{2} \left\Vert \frac{\mathbf{h}_i \ast \mathbf{x}_i - \mathbf{y}_i}{\sigma_i} \right\Vert_F^2  + \lambda_{i} \textbf{K} \label{iloss}\\[10pt]
    L_z(\mathbf{x}_z) &= \frac{1}{2} \left\Vert \frac{\mathbf{h}_z \ast \mathbf{x}_z - \mathbf{y}_z}{\sigma_z} \right\Vert_F^2  + \lambda_{z} \textbf{K,} \label{zloss}
\end{align}

\begin{align}
\text{where } \textbf{K} = \left\Vert \frac{\hspace{2pt} \mathbf{h}_{euc} \ast \sum\limits_{c\in\{r,i,z\}} \alpha_c \mathbf{x}_c - \mathbf{y}_{euc} }{\sigma_{euc}} \right\Vert_F^2 \label{eq:constr}.
\end{align}

\noindent The first terms in Eqs. \ref{vloss}-\ref{zloss} represent the data fidelity terms for each respective band, with $\sigma_r$, $\sigma_i$, and $\sigma_z$ being the corresponding noise maps. The second terms are the constraining terms that enforce the condition that the sum of the Rubin $r$-, $i$-, and $z$-band images equals the Euclid VIS-band image, as expected from Fig. \ref{fig:filters} after flux calibration. Within these constraining terms, the individual images $\mathbf{x}_r$, $\mathbf{x}_i$, and $\mathbf{x}_z$ are multiplied by their respective fractional flux contributions $\alpha_r$, $\alpha_i$, and $\alpha_z$, which represent the fractional area overlaps between their corresponding filter curves. The values of $\alpha_r$, $\alpha_i$, and $\alpha_z$ are obtained by integrating the area under the curves in Fig. \ref{fig:filters} and normalizing them to sum up to one. The resulting values are $\alpha_r=0.3785$, $\alpha_i=0.3468$, and $\alpha_z=0.2746$. The denominator, $\sigma_{euc}$, denotes the Euclid image noise map. The choice of the multiplicative hyperparameters $\lambda_{r}$, $\lambda_{i}$, and $\lambda_{z}$ is described in Sect. \ref{hyperparam}.

\subsection{Hyper-parameter tuning}
\label{hyperparam}
The hyper-parameters $\lambda_{r}$, $\lambda_{i}$, and $\lambda_{z}$ are determined by varying the ratios between the second and first terms in Eqs. \ref{vloss}-\ref{zloss}. These ratios indicate the contribution coming from the constraining term. We varied the ratios from 0 (no contribution from the constraining term) to 1 (equal contribution from the constraining term) in steps of 0.01. From this experiment, the optimal solution yields the lowest mean squared error with a ratio of 0.3 for all three photometric bands. Subsequently, the values of $\lambda_{r}$, $\lambda_{i}$, and $\lambda_{z}$ are computed from these ratios by dividing them by $\textbf{K}$ (Eq. \ref{eq:constr}).

\subsection{Optimization}
\label{optimization}
The aim is to find optimal solutions $\mathbf{\hat{x}}_r$, $\mathbf{\hat{x}}_i$, $\mathbf{\hat{x}}_z$ that minimize the individual loss functions:
\begin{align*}
    \mathbf{\hat{x}}_{\{r,i,z\}} = \underset{\mathbf{x}_{\{r,i,z\}}}{\operatorname{argmin}} \hspace{2pt}L_{\{r,i,z\}}(\mathbf{x}_{\{r,i,z\}}),
\end{align*}

\noindent which is done in an alternative and iterative manner using gradient descent:
\begin{align}
    \mathbf{x}_{\{r,i,z\}}^{[k+1]} = \mathbf{x}_{\{r,i,z\}}^{[k]} - \beta_{\{r,i,z\}} \nabla L_{\{r,i,z\}}(\mathbf{x}_{\{r,i,z\}}^{[k]}), \label{eq:graddesc}
\end{align}

\noindent where $\mathbf{x}^{[k]}$ denotes the variable $\mathbf{x}$ at $k^{th}$ iteration, and $\beta_r, \beta_i, \beta_z \in \mathbb{R}^{n}$ are the step sizes chosen such that convergence is guaranteed (described in more detail in Sect. \ref{stepsize}). While computing Eq. \ref{eq:graddesc} for one band, it is assumed that the other two bands are known and remain constant. The gradients of the loss functions \ref{vloss}-\ref{zloss} are given by

\begin{align}
        \nabla L_r(\mathbf{x}_r) &= \frac{\mathbf{h}_r^\top \ast (\mathbf{h}_r \ast \mathbf{x}_r - \mathbf{y}_r)}{\left\Vert \mathbf{\sigma}_r \right\Vert_F^2} + \lambda_{r} \alpha_r \mathbf{K}_{grad} \\[5pt]
        \nabla L_i(\mathbf{x}_i) &= \frac{\mathbf{h}_i^\top \ast (\mathbf{h}_i \ast \mathbf{x}_i - \mathbf{y}_i)}{\left\Vert \mathbf{\sigma}_i \right\Vert_F^2} + \lambda_{i} \alpha_i \mathbf{K}_{grad} \\[5pt]
        \nabla L_z(\mathbf{x}_z) &= \frac{\mathbf{h}_z^\top \ast (\mathbf{h}_z \ast \mathbf{x}_z - \mathbf{y}_z)}{\left\Vert \mathbf{\sigma}_z \right\Vert_F^2} + \lambda_{z} \alpha_z \mathbf{K}_{grad},
\end{align}

\begin{align}
    \text{where } \textbf{K}_{grad} =  \frac{2 \mathbf{h}_{euc}^\top}{\left\Vert \mathbf{\sigma}_{euc} \right\Vert_F^2} \ast \left( \hspace{2pt} \mathbf{h}_{euc} \ast \sum\limits_{c\in\{r,i,z\}} \alpha_c \mathbf{x}_c - \mathbf{y}_{euc}\right).
\end{align}

\subsection{Gradient descent step size}
\label{stepsize}
Suppose a function $f: \mathbb{R}^{n\times n} \xrightarrow{} \mathbb{R}^{n\times n}$ that is convex and differentiable. Its gradient is Lipschitz-continuous if there exists some constant C such that
\begin{align*}
    \left\Vert \nabla f(\mathbf{x'}) - \nabla f(\mathbf{x}) \right\Vert &\leq C \left\Vert \mathbf{x'} - \mathbf{x} \right\Vert.
\end{align*}

Since the loss functions \ref{vloss}-\ref{zloss} are convex and differentiable, one could find Lipschitz constants $C_{\{r,i,z\}}$ such that
\begin{align*}
    &\left\Vert \nabla L_{\{r,i,z\}}(\mathbf{x'}_{\{r,i,z\}}) - \nabla L_{\{r,i,z\}}(\mathbf{x}_{\{r,i,z\}}) \right\Vert \leq C_{\{r,i,z\}} \left\Vert \mathbf{x'}_{\{r,i,z\}} - \mathbf{x}_{\{r,i,z\}} \right\Vert
\end{align*}
\begin{align}
    &C_{\{r,i,z\}} \geq \frac{\mathbf{h}^\top_{\{r,i,z\}} \ast \mathbf{h}_{\{r,i,z\}}}{\left\Vert \mathbf{\sigma}_{\{r,i,z\}} \right\Vert_F^2} + 2 \lambda_{\{r,i,z\}} \alpha^2_{\{r,i,z\}} \mathbf{h}_{euc}^\top \ast \mathbf{h}_{euc}. \label{eq:lipschitz}
\end{align}

\noindent Once that is found, the optimal constraints on the individual step sizes, $\beta_{\{r,i,z\}}$, that ensure convergence are as follows:
\begin{align}
    \beta_{\{r,i,z\}} &\leq \frac{1}{C_{\{r,i,z\}.}} \label{eq:stepsize}
\end{align}

\noindent From Eqs. \ref{eq:lipschitz} and \ref{eq:stepsize}, it is important to note that the step size also depends on the Rubin and Euclid PSFs (see details in Sects. \ref{sec:lsst_gen} and \ref{sec:euclid_gen}) and the values of $\lambda_{\{r,i,z\}}$ (described in Sect. \ref{hyperparam}) and $\alpha_{\{r,i,z\}}$ (shown in Sect. \ref{forward_model}).

\section{Dataset generation}
\label{sec:data_exp}

\subsection{Ground truth images}
\label{sec:gt_hst}

We extracted HST cutout windows of dimensions $128 \times 128$ pixels from GOODS-N and GOODS-S \citep{goods} in the $F606W$, $F775W$, and $F850LP$ bands by centering them at the centroid of the object. These HST bands were selected because their central wavelengths align with those of the Rubin $r$, $i$, and $z$ bands, and these HST images were subsequently used to simulate the Rubin images, as explained in Sect. \ref{sec:lsst_gen}. The mosaicked HST ACS (Advanced Camera for Surveys) images along with the catalog can be found at this link\footnote{\url{https://archive.stsci.edu/prepds/goods/}\label{foot}}. These HST cutouts are at a pixel scale of $0.05\arcsec$. We aimed to perform the experiments on galaxies with sizes large enough to effectively assess the impact of deconvolution on our ability to resolve their structural and morphological features, such as arms, bars, and clumps. To ensure the selection of large galaxies and exclude point-sized objects, we applied the following filtering criteria to the $F775W$ band catalog:
\begin{itemize}
    \item $18 <$ MAG\_AUTO $< 23$  (AB magnitude in SExtractor “AUTO” aperture)
    \item $\text{Flux\_Radius}_{80} > 10$ ($80\%$ enclosed flux radius in pixels)
    \item FWHM $> 10$ (full width at half maximum in pixels)
\end{itemize}

We then visually inspected and selected 92 objects that exhibited extended and complex galaxy structures. The histogram of the HST $F775W$ band magnitude for all galaxies in our dataset is shown in Fig. \ref{fig:mag_hist}. It is important to highlight that, according to our simulations, the HST $F775W$ band corresponds to the Rubin $i$ band, and we refer to it as the Rubin $i$ band throughout the text.

\begin{figure}[h!]
    \centering
    \label{fig:mag_hist}\includegraphics[width=0.95\linewidth]{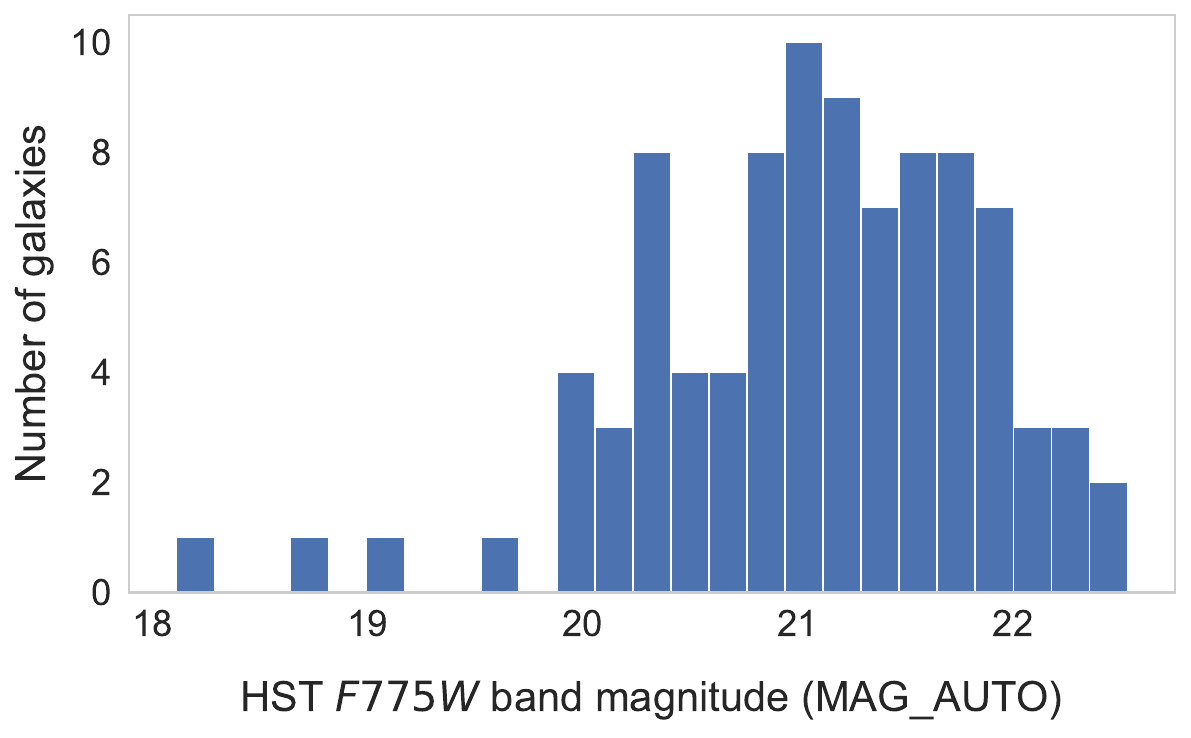}
    \caption{Histogram of the HST $F775W$-band magnitude for all galaxies in our dataset after filtering. Note that the HST $F775W$ band matches with the Rubin $i$ band.}
    \label{fig:mag_hist}
\end{figure}

\subsection{Vera C. Rubin images}
\label{sec:lsst_gen}

The simulated Rubin-like PSFs were obtained from the second data challenge (DC2) of the Legacy Survey of Space and Time (LSST) Dark Energy Science Collaboration (DESC). The atmospheric and optical effects, as well as sensor-induced electrostatic effects, are simulated using physically motivated models, with a final adjustment to the PSF sizes to match the expected data from the Rubin Observatory, as described in detail in \cite{lsst_2021}.

To generate the Rubin simulated images, we first convolved the HST images in the $F606W$, $F775W$, and $F850LP$ bands with the Rubin $r$-, $i$-, and $z$-band PSFs, respectively, such that the simulated images are at the expected Rubin resolution, with a pixel scale of $0.2\arcsec$. Subsequently, we added white Gaussian noise such that our Rubin-simulated images have a S/N ranging between $12$ and $28$, with a median around 20. Based on the survey parameters outlined in \cite{lsst_2019}, our simulations suggest an S/N range that corresponds to the initial few visits of the telescope. This indicates that our method could be effectively applied as soon as the first images start arriving.

\subsection{Euclid images}
\label{sec:euclid_gen}

The simulated Euclid-like VIS-band PSF was obtained using the WaveDiff model proposed by \cite{tobias2022}, which changes the data-driven PSF modeling
space from the pixels to the wavefront by adding a differentiable optical forward model in the modeling framework. WaveDiff outputs an approximation of the true Euclid PSF, which was derived before the actual launch of the satellite. 

Next, we calculated the fractional flux contributions $\alpha_r, \alpha_i, \alpha_z$ by integrating the area under the curves in Fig. \ref{fig:filters} and normalizing them to sum to 1. The resulting values, as also mentioned in Sect. \ref{forward_model}, are $\alpha_r=0.3785$, $\alpha_i=0.3468$, and $\alpha_z=0.2746$. To generate the Euclid simulated images, we multiplied $\alpha_r, \alpha_i$, and  $\alpha_z$ by the HST images in the $F606W$, $F775W$, and $F850LP$ bands, respectively. The result is then convolved with the simulated Euclid PSF such that the simulated images are at the expected Euclid resolution with a pixel scale of $0.1\arcsec$. Finally, white Gaussian noise is added such that the Euclid-simulated images have a S/N ranging between $20$ and $45$, with a median around 35. Based on the calculations presented by \cite{Euclid2}, who assessed the S/N statistics for Euclid, our simulations are conservative, implying that our method would perform well when applied to real images with higher S/Ns.

\section{Results}
\label{results}

The algorithm simultaneously processed the noisy simulations from the three Rubin bands and the Euclid VIS band, along with their respective PSFs. These noisy images served as initializations or first guesses for the algorithm. Subsequently, the algorithm iteratively minimized the loss functions \ref{vloss}-\ref{zloss}, as detailed in Sect. \ref{optimization}. The algorithm was run for 200 iterations, with convergence typically observed within $50$-$100$ iterations for all images in our dataset. Figure \ref{fig:loss} shows the convergence plot of the loss function for the deconvolved output in Fig. \ref{subfig:mcdec1}.

\begin{figure}[h!]
    \centering
    \label{fig:loss}\includegraphics[width=0.95\linewidth]{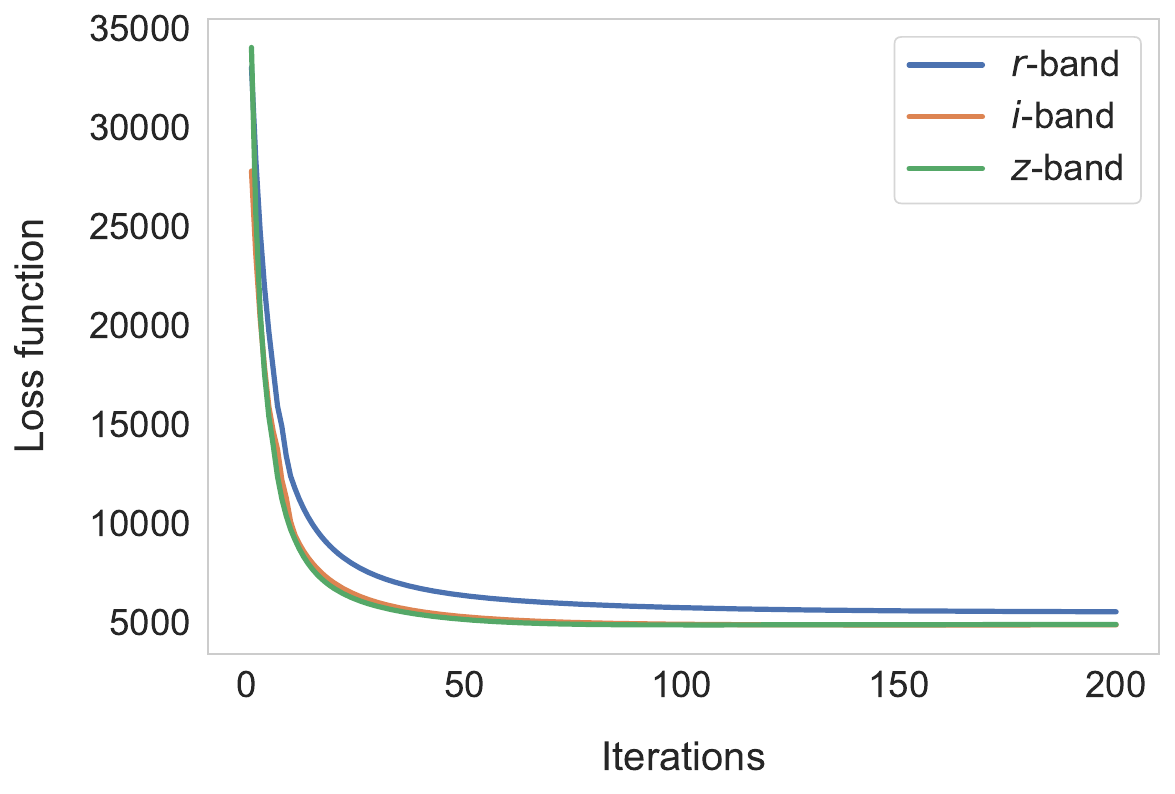}
    \caption{Loss function for the galaxy shown in Fig. \ref{subfig:mcdec1}. Convergence is guaranteed at around 100 iterations when the relative change in loss value is $<10^{-3}$ and the curve is flat.}
    \label{fig:loss}
\end{figure}

\subsection{Flux leakage test}
\label{flux_leakage}

\begin{figure}[h!]
    \centering
    \label{subfig:flux_leakage}\includegraphics[width=\linewidth]{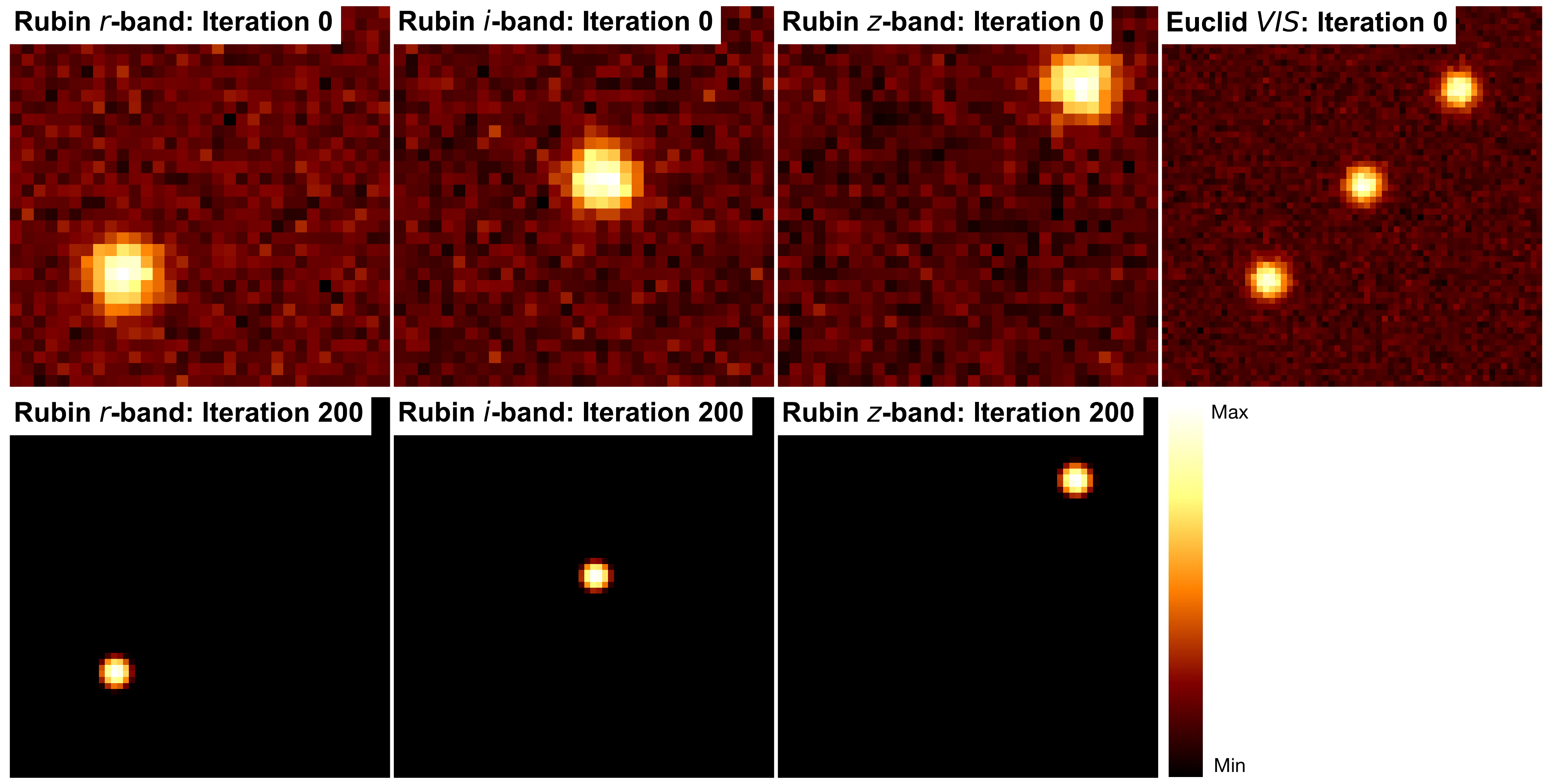}

    \caption{Unit test to verify that there is no leakage of flux from one channel to another. The recovered Gaussians remain at their original centers.}
    \label{fig:flux_leakage}
\end{figure}

As a validation to ensure no flux leakage between channels during the joint deconvolution of Rubin and Euclid images, we conducted a unit test. We assumed three distinct Gaussians placed separately in the Rubin channels, with the Euclid simulation being a weighted sum of these three images, resulting in three disjoint Gaussians. Post-deconvolution, the Gaussians remained intact without any structure extending beyond their boundaries. This confirms that the structure present in the deconvolved image within each Rubin band is independent of the structures in other bands and each image accurately retains only the information relevant to its specific band. Figure \ref{fig:flux_leakage} provides a visual demonstration of these findings. 

Moreover, to further validate that our method is effective for objects with non-flat spectral energy distributions (SEDs), we analyzed the transfer of information across different bands for such objects. The results are presented in Appendix \ref{sec:noise_test}.

\subsection{Deconvolved outputs}
\label{sec:deconv}

We present two examples of deconvolved images in Fig. \ref{fig:deconv_outputs}, illustrating the algorithm's capability to recover features that were lost in the original Rubin simulations. Visually, the deconvolved outputs seem to capture the variations between different bands. Qualitatively, these outputs exhibit high quality with minimal background noise and result in clean residuals. However, in Sect. \ref{sec:denoise}, we demonstrate that employing a deep-learning-based denoiser further enhances the already impressive results achieved initially. Table \ref{tab:clusters} presents the NMSE computed with respect to the ground-truth HST image for the pre-denoised and post-denoised images.

\subsection{Deep-learning-based denoising}
\label{sec:denoise}

\begin{figure*}[h!]
\begin{center}
\includegraphics[width=\textwidth]{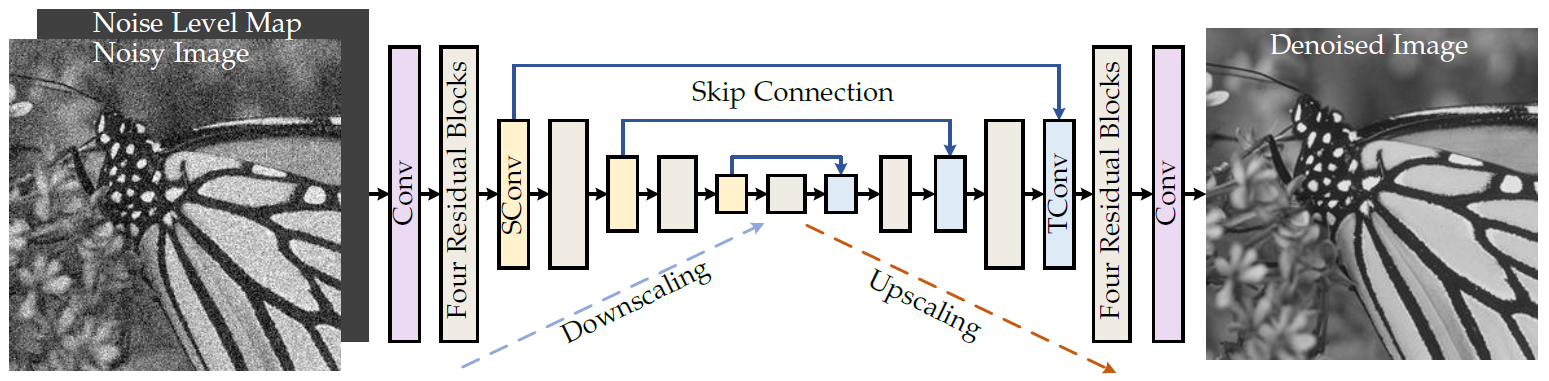}
\end{center}
\caption{\label{fig:flux_preserve} {DRUNet architecture, which incorporates an additional noise level map as input and integrates U-Net \citep{ronnenberger2015unet} with ResNet \citep{resnet}. "SConv" stands for strided convolution, and "TConv" stands for transposed convolution. Image credits: \cite{drunet}.}}\label{fig:drunet}
\end{figure*}

After obtaining the deconvolved outputs, we feed them to DRUNet, a neural network proposed by \cite{drunet} that combines U-Net \citep{ronnenberger2015unet} and ResNet \citep{resnet}. U-Net  is renowned for its efficiency in image-to-image translation, while ResNet excels in increasing modeling capacity through stacked residual blocks. Inspired by FFDNet \citep{ffdnet}, which incorporates a noise level map as input, DRUNet enhances U-Net by integrating residual blocks to improve prior denoising modeling. Similar approaches that combine U-Net and ResNet can be found in other studies \citep{road_ext, VenkateshG2018ADR}. The backbone of DRUNet is a U-Net architecture with four scales, and the schematic proposed by \cite{drunet} is depicted in Fig. \ref{fig:drunet}. 

\begin{figure*}[h!]
\centering
    \subfigure[\makebox{}]
    {\label{subfig:druunet1}\includegraphics[width=0.93\linewidth]{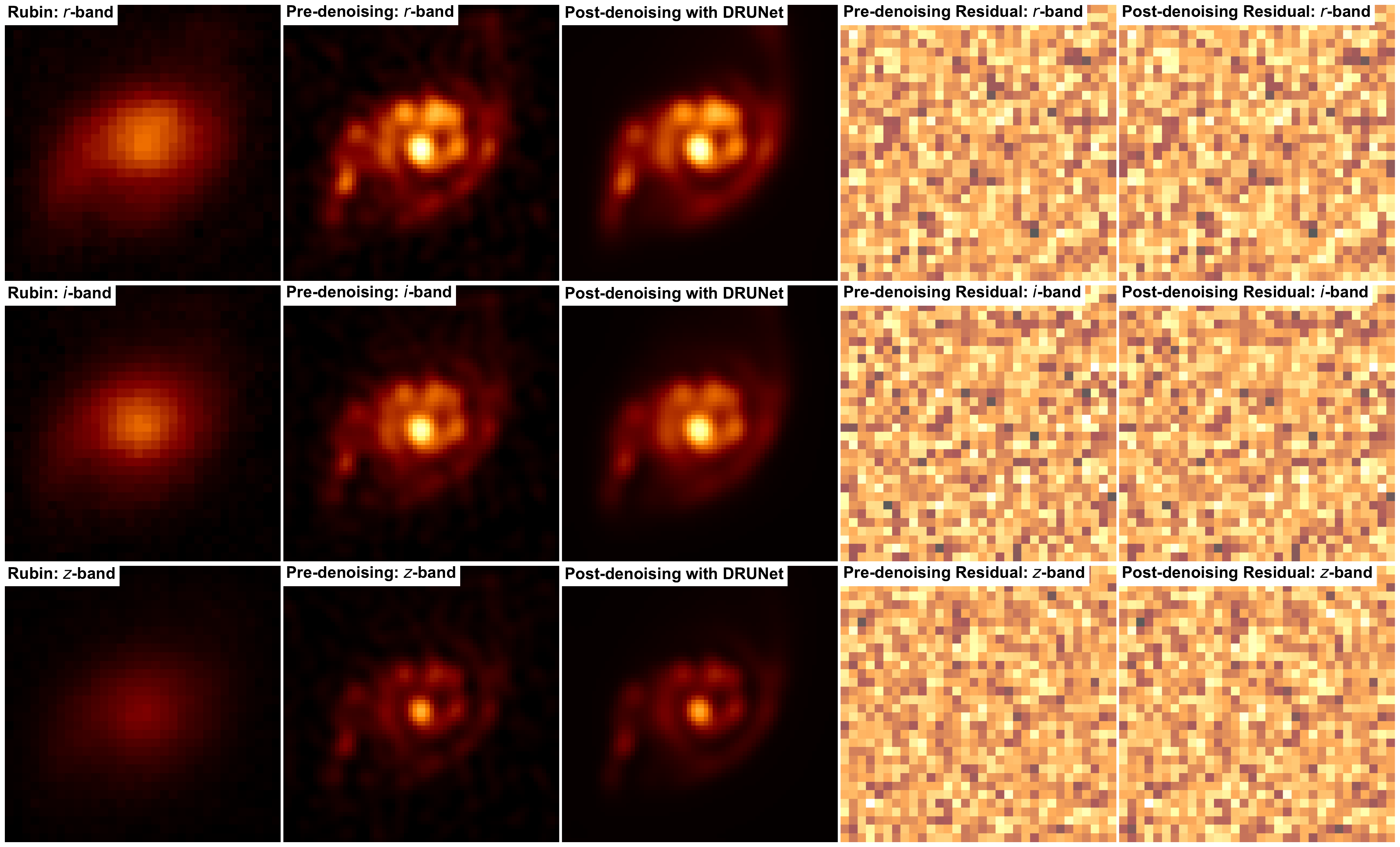}}\\
    
    \subfigure[\makebox{}]
    {\label{subfig:druunet2}\includegraphics[width=0.93\linewidth]{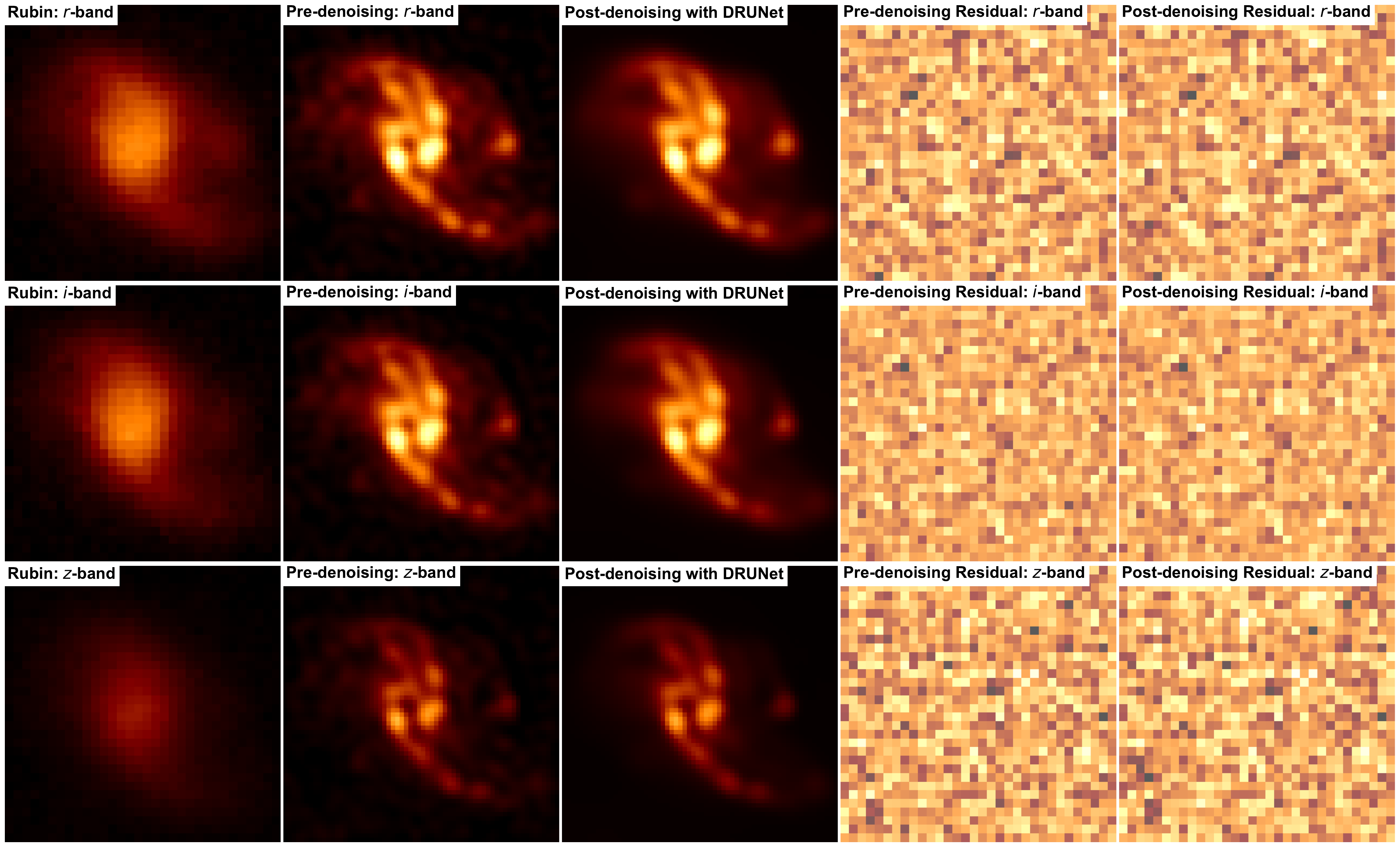}}
\caption{\label{fig:denoise_outputs} {Two denoised outputs from DRUNet that demonstrate its ability to effectively remove noise from the background while preserving the structure of the central galaxy. First column: Rubin simulations in the $r$, $i$, and $z$ bands. Second: Corresponding pre-denoised outputs. Third: Corresponding post-denoised outputs. Fourth and fifth: Residuals.}}
\end{figure*}

We chose DRUNet because it is a non-blind denoiser, meaning it takes the noise map as input. This ensures that any unknown noise can be estimated and given as input for denoising. We estimate the noise level map in our deconvolved images using the scikit-image package \citep{skimage} by calculating the average noise level within four $16 \times 16$ pixel squares placed at each corner of the image. This approach ensures that only background noise is measured, avoiding any contribution from the signal. This map is then fed to the pre-trained DRUNet, along with the deconvolved image. The denoised outputs are illustrated in Fig. \ref{fig:denoise_outputs}. It is observed that the denoiser exclusively eliminates noise from the background without affecting the main structure of the galaxies. Although the enhancement in image quality is marginal (since the original deconvolved image is already of high quality), the NMSE computed with respect to the ground-truth HST image decreases, as shown in Table \ref{tab:clusters}. The most notable improvement is observed in the $z$-band images. Even though DRUNet was trained on a combination of images from the BSD \citep{bsd}, Waterloo Exploration Database \citep{waterloo}, DIV2K \citep{div2k}, and Flick2K \citep{flick}, it is remarkable that it works well on astronomical data, showing great generalization. Finally, the fractional error in the output flux as a function of the $i$-band magnitude is shown in Fig. \ref{fig:flux_preserve}, which indicates that the mean flux error is less than $5\%$ for the entire magnitude range for all the bands.

\begin{figure*}[h!]
\begin{center}
\includegraphics[width=\textwidth]{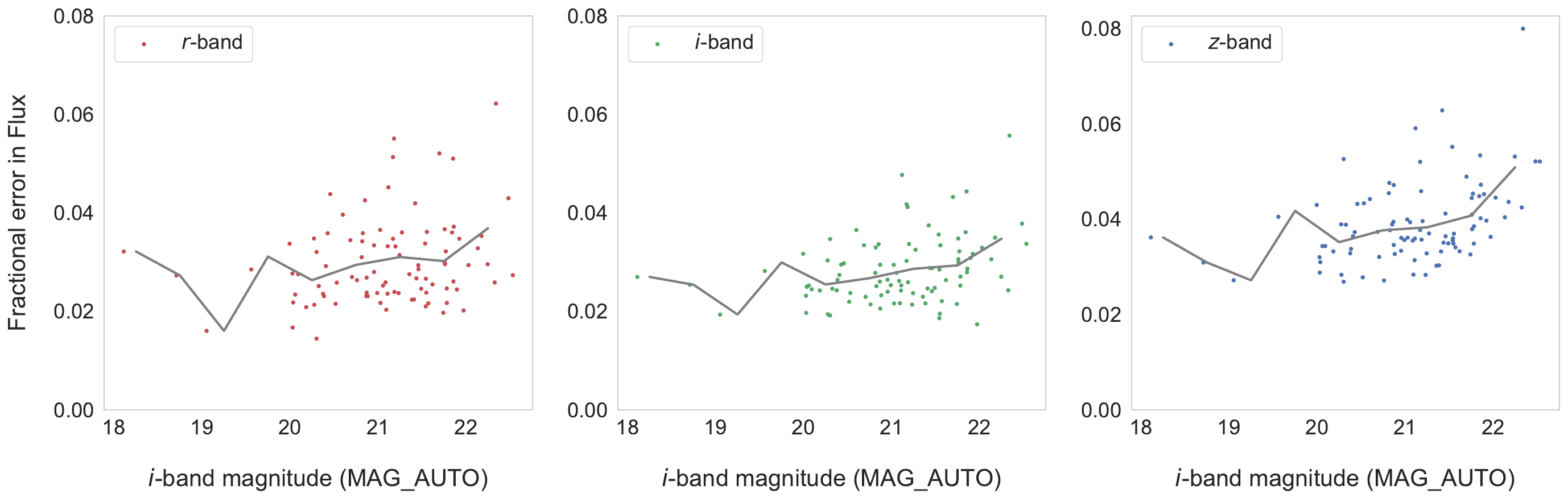}
\end{center}
\caption{\label{fig:flux_preserve} {Fractional error in the output flux as a function of the $i$-band magnitude (which is chosen in order to have the same scale on the x-axes). The dots correspond to the individual galaxies, and the gray line is the best-fit line after binning the magnitude values.}}
\end{figure*}

\begin{table}[h!]
    \centering
    \renewcommand{\arraystretch}{1.3}
        \caption{NMSE with respect to HST images.}
    \begin{tabular}{cccc}
        \hline
        \textbf{NMSE} & \textbf{$r $ band} & \textbf{$i$ band} & \textbf{$z$ band}\\
        \hline
        Pre-denoising & 0.059 & 0.041 & 0.053\\
        Post-denoising & 0.058 & 0.038 & 0.038\\
        \hline
        \% improvement & 1.69\% & 7.32\% & 28.3\%\\
        \hline
    \end{tabular}
    \tablefoot{The NMSE is calculated for the pre-denoised and post-denoised images with respect to the ground-truth HST image.}
    \label{tab:clusters}
\end{table}

\subsection{Improvement compared to an independent deconvolution of each band}
\label{sec:sunet_deconv} 

To demonstrate the advantages of jointly deconvolving multiple photometric bands, we performed independent deconvolution of each band using the deconvolution framework introduced in \cite{sunet_akhaury}. The method proceeds in two steps: it first de-convolves the input using Tikhonov regularization \citep{tikhonov1977solutions}, and denoises the result using SUNet \citep{sunet}, a state-of-the-art Swin transformer-based architecture. The results, presented in Fig. \ref{fig:deconv_comp}, show that the joint deconvolution method outperforms the independent deconvolution of individual bands. The joint method enables us to leverage the correlation between the different bands and the space-based image, thus improving the final output.

\section{Conclusion}
\label{concl}

We have presented a novel image deconvolution technique designed to improve the resolution of multiband ground-based data by leveraging higher-resolution space-based observations. Our approach, which focuses on the joint deconvolution of Rubin and Euclid images, effectively exploits the overlapping spectral coverage of the Rubin $r$, $i$, and $z$ bands with the Euclid VIS band. Through rigorous testing, we have demonstrated that our iterative algorithm successfully recovers fine details while preserving the flux for each band. Different noise levels were tested, and the resolution achieved by ground-based data is close to that of HST. Our results indicate that a joint deconvolution of all data outperforms independent deconvolutions of individual photometric bands using existing state-of-the-art methods. By utilizing the Euclid VIS-band image as a term that provides additional information, our technique ensures that the deconvolved Rubin images retain high spatial resolution and accurate photometric measurements. The integration of deep-learning-based denoising using DRUNet enhances the quality of the deconvolved output, further reducing background noise without altering the main structures of the galaxies.

The potential applications of our method extend beyond the Euclid-Rubin pair, offering a versatile solution to improving the resolution of ground-based images in multiple photometric bands as long as there exists a space-based image of the same field of view in a band that encompasses all ground-based filters. In the future, we intend to test our deconvolution method on images of the Perseus cluster by using ground-based observations from the Canada–France–Hawaii Telescope (CFHT) and space-based observations from the Euclid Early Release Observations (ERO) public release.

\section*{Data availability}
\label{sec:reproducible_research}

For the sake of reproducible research, the codes used for this article are publicly available online. 

\begin{enumerate}
    \item The ready-to-use version of our joint deconvolution method\footnote{\url{https://github.com/utsav-akhaury/Multiband-Deconvolution/tree/main}}.
    \item The DRUNet repository from \cite{drunet} \footnote{\url{https://github.com/cszn/DPIR}}.
\end{enumerate}

\begin{acknowledgements}
      This work was funded by the Swiss National Science Foundation (SNSF) under the Sinergia grant number CRSII5\_198674. This work was supported by the TITAN ERA Chair project (contract no. 101086741) within the Horizon Europe Framework Program of the European Commission, and the  Agence Nationale de la Recherche (ANR-22-CE31-0014-01 TOSCA).
\end{acknowledgements}

\begin{figure*}[h!]
\centering
    \subfigure[\makebox{}]
    {\label{subfig:mcdec1}\includegraphics[width=0.95\linewidth]{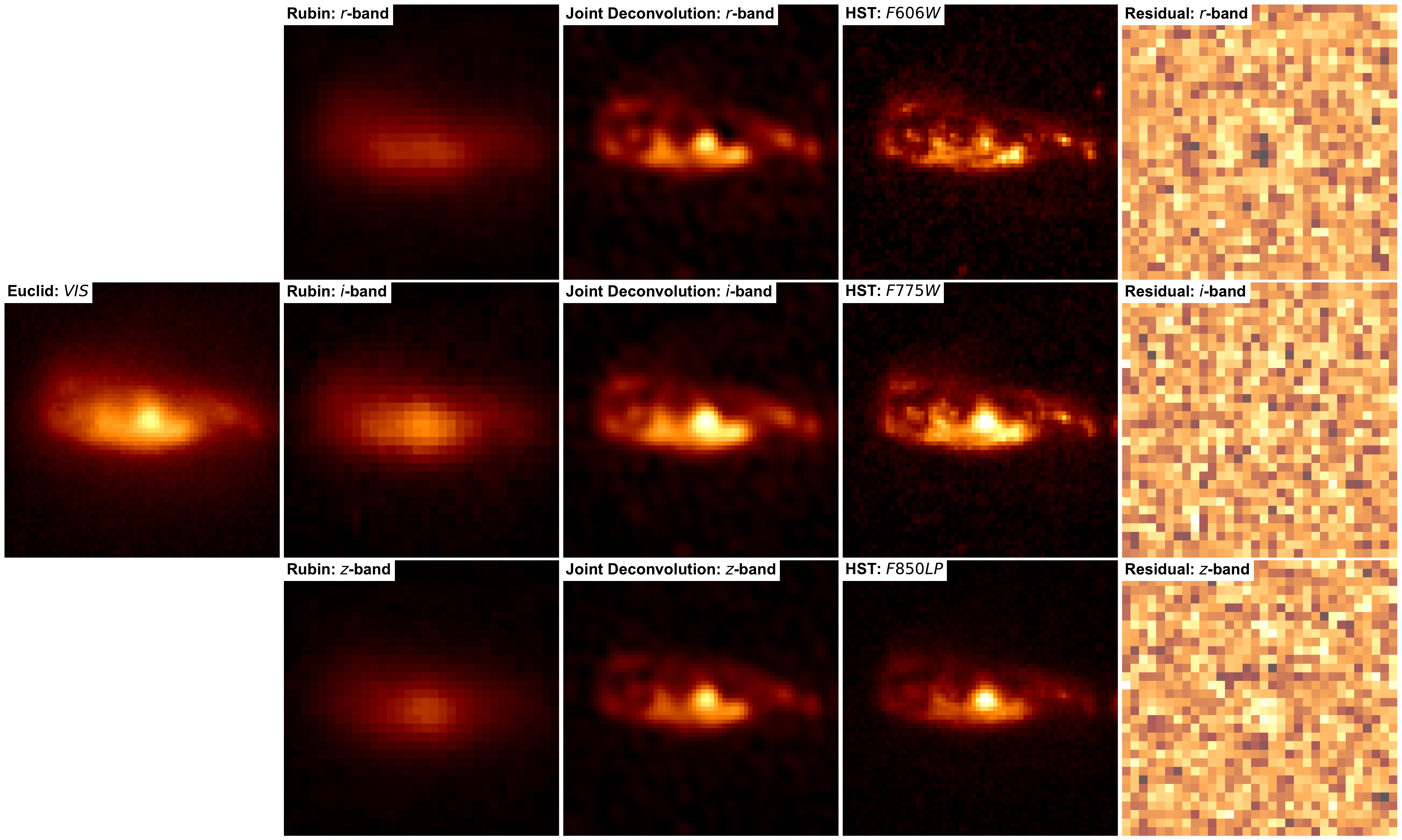}}\\
    
    \subfigure[\makebox{}]
    {\label{subfig:mcdec2}\includegraphics[width=0.95\linewidth]{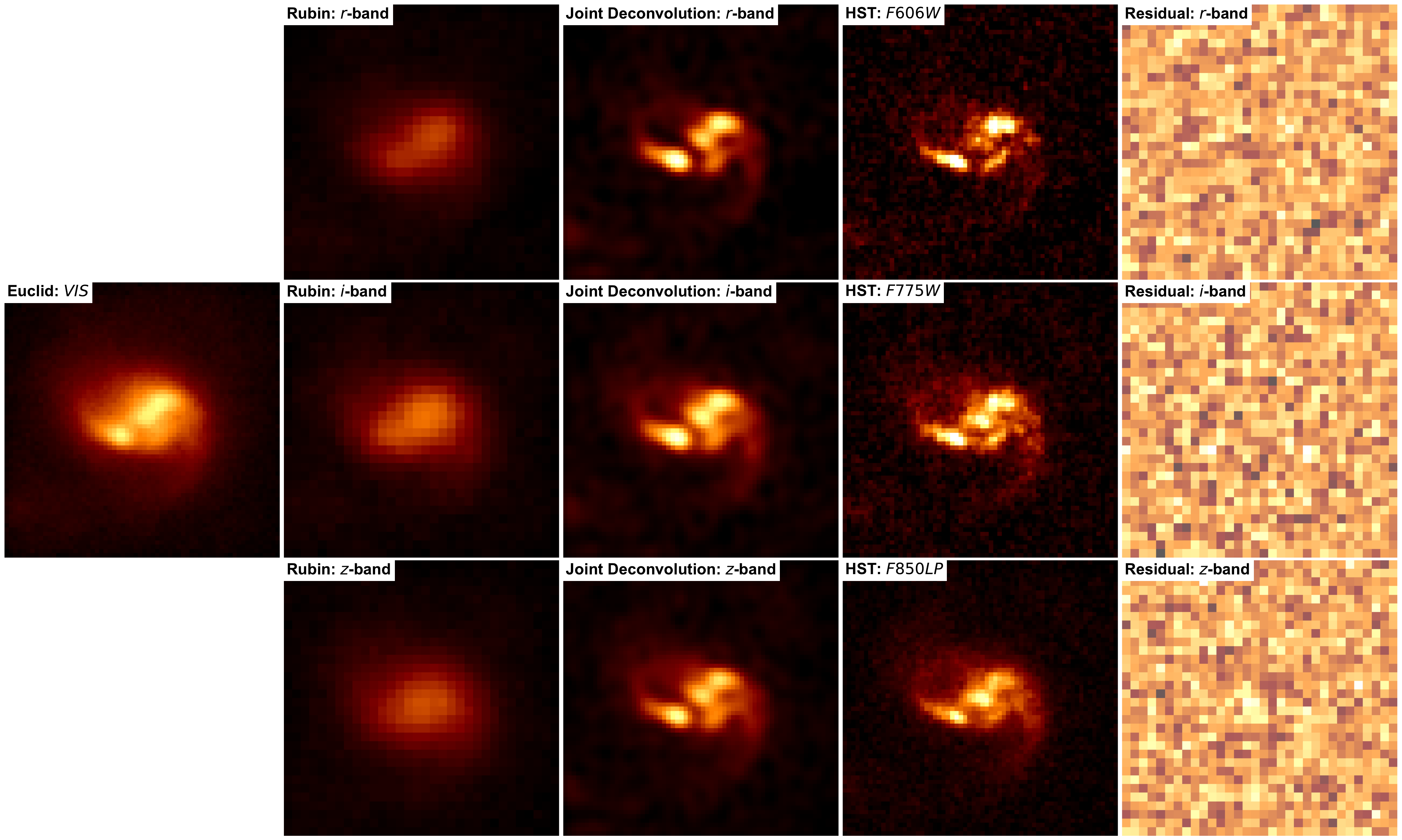}}
\caption{\label{fig:deconv_outputs} {Two deconvolved outputs that illustrate the successful recovery of features completely lost in the Rubin simulations. Additionally, the outputs seem to capture the variations when transitioning from one band to another. First column: Euclid VIS image.\ Second: Rubin simulations in the $r$, $i$, and $z$ bands. Third: Deconvolved outputs for the three bands. Fourth: Corresponding ground-truth HST images.\ Fifth: Residuals, which are defined as follows: residual $=$ noisy Rubin image $-$ PSF $\ast$ deconvolved image.}}
\end{figure*}

\begin{figure*}[h!]
\centering
    \includegraphics[width=0.95\linewidth]{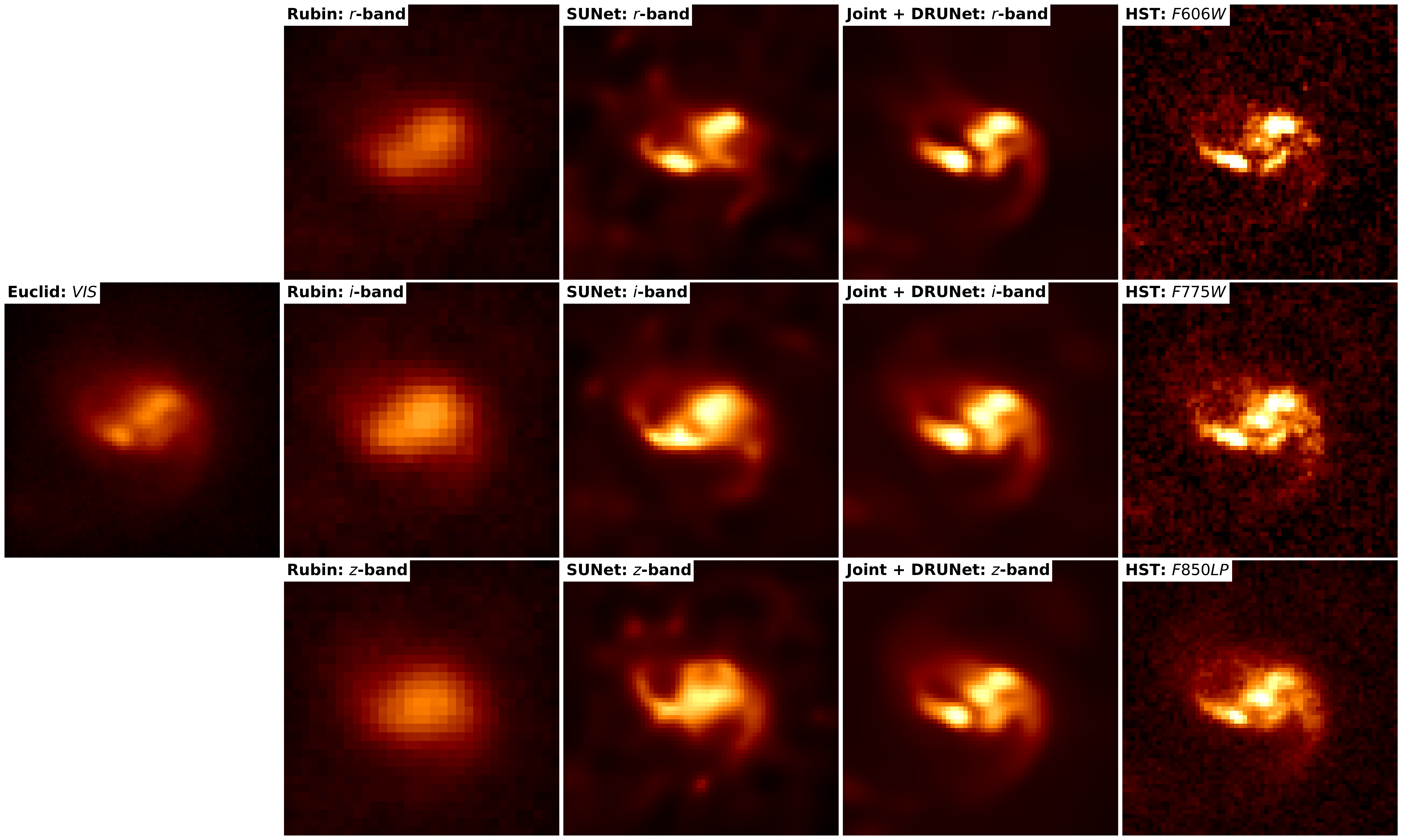}
\caption{\label{fig:deconv_comp} {Galaxy shown in Fig. \ref{subfig:mcdec2} deconvolved using two different approaches, illustrating that joint deconvolution outperforms independent deconvolutions of individual photometric bands. The joint method allows us to leverage the correlation between the different bands and the space-based image, thus improving the final output. First column: Euclid VIS image. Second: Rubin simulations in the $r$, $i$, and $z$ bands. Third: Independently deconvolved SUNet outputs for the three bands. Fourth: Corresponding joint deconvolution outputs followed by denoising with DRUNet. Fifth: Ground-truth HST images.}}
\end{figure*}

\bibliographystyle{aa}
\bibliography{references.bib}

\begin{appendix} 

\section{Generalization to objects with non-flat SEDs}
\label{sec:noise_test}

\begin{figure}[h!]
    \centering

    \subfigure[\makebox{}]
    {\label{subfig:noise_test1}\includegraphics[width=\linewidth]{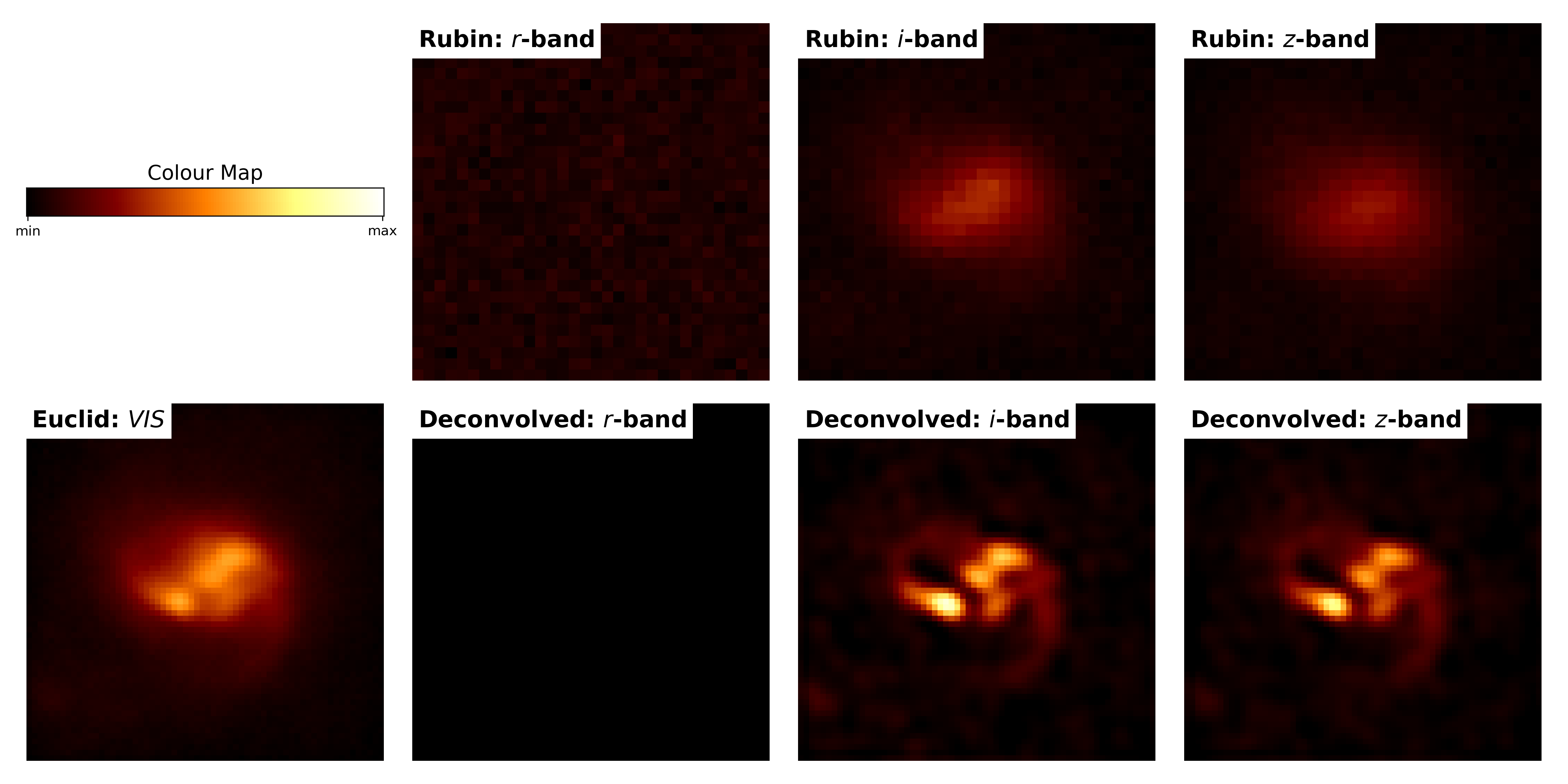}}\\

    \subfigure[\makebox{}]
    {\label{subfig:noise_test2}\includegraphics[width=\linewidth]{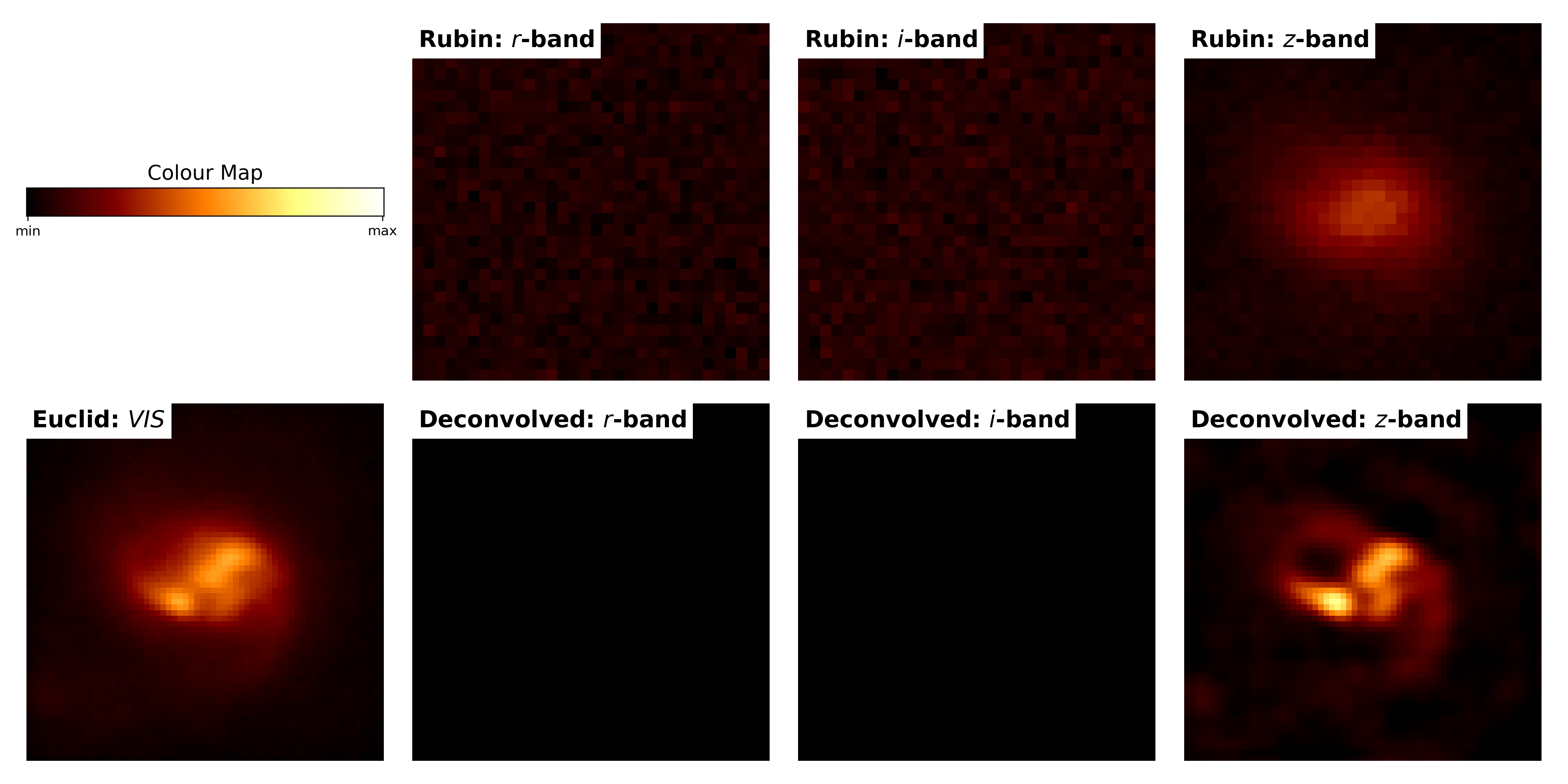}}\\

    \subfigure[\makebox{}]
    {\label{subfig:noise_test3}\includegraphics[width=\linewidth]{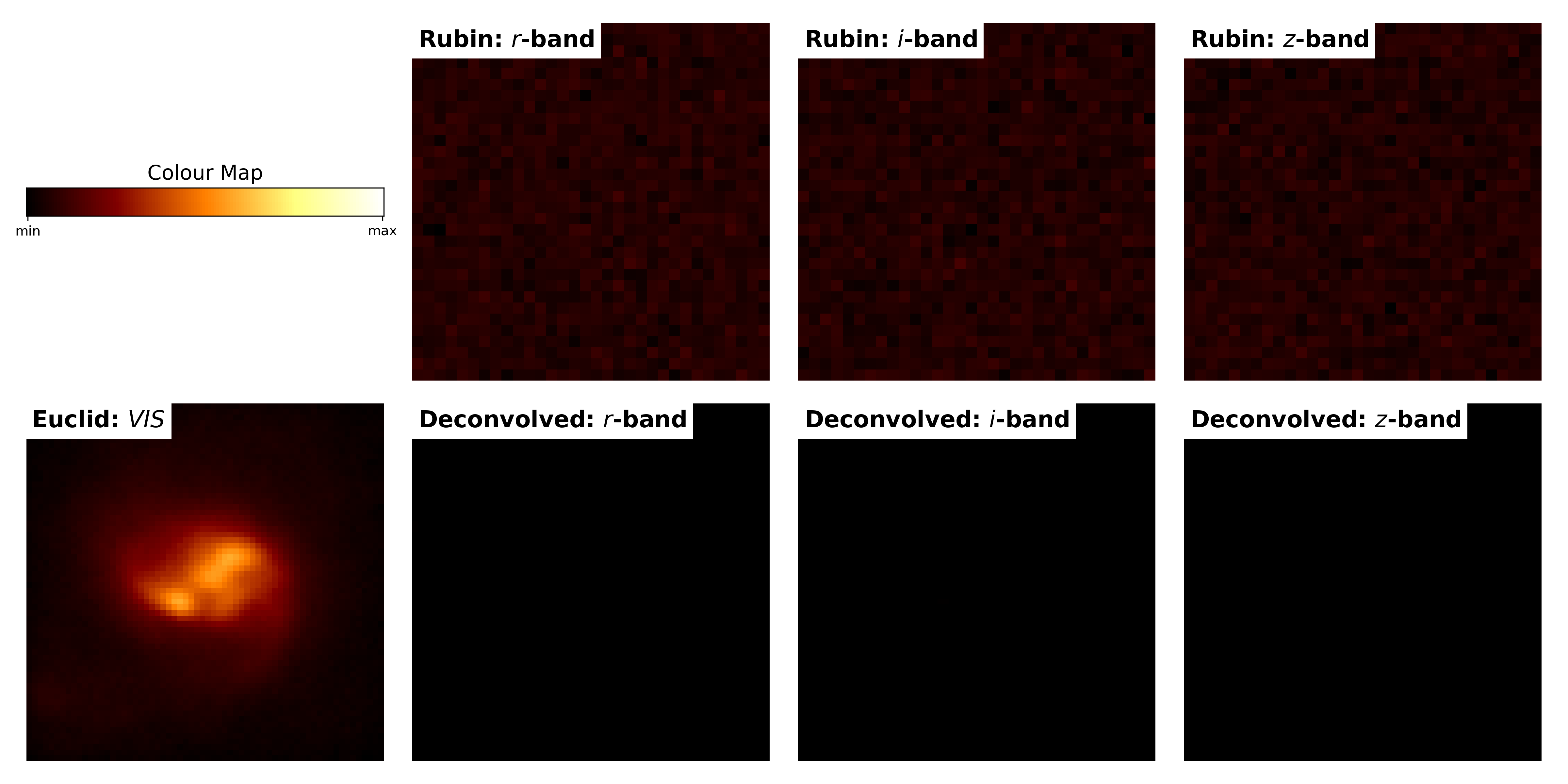}}\\
    
    \caption{Galaxy from Fig. \ref{subfig:mcdec2}, with each band successively replaced by a random noise map, as shown when progressing from Figs. \ref{subfig:noise_test1} to \ref{subfig:noise_test3}. Reconstructed features appear only in bands where the original galaxy signal is present.}
    \label{fig:noise_test}
\end{figure}

To assess the impact of information transfer across different bands for objects with a non-flat SED, we conducted an experiment where we successively replaced the Rubin bands with pure noise, ensuring that no galaxy signal was present. We then analyzed how features from the high-resolution Euclid image propagated into the deconvolved Rubin bands. The outputs are shown in Fig. \ref{fig:noise_test}.

Our results confirm that reconstructed features appear only in bands where the original galaxy signal is present. This demonstrates that the algorithm does not artificially imprint Euclid information onto Rubin bands lacking real data. In cases where a structure is visible in the Euclid image but absent from one or more Rubin bands, the problem becomes degenerate: the feature could either be entirely attributed to a single band or distributed across multiple bands. However, our findings indicate that the outputs are directly influenced by the input data in each band rather than being dictated solely by the high-resolution Euclid image. Our results demonstrate that the joint deconvolution method effectively utilizes the available signal in each band while respecting the constraints imposed by the data. This also confirms that the method would work for objects with non-flat SEDs, where signal may be present in only one band but absent in others, ensuring that features are accurately transferred according to their actual distribution across the bands. 

As mentioned in Sect. \ref{results}, we find that the algorithm converges within $200$ iterations when the signal is available in all three bands. When one band is replaced with a noise map, as shown in Fig. \ref{subfig:noise_test1}, convergence requires approximately $1000$ iterations. Replacing a second band with a noise map, as seen in Fig. \ref{subfig:noise_test2}, further increases the iteration count to around $5000$. This demonstrates that incorporating more data across different bands significantly accelerates the convergence of the loss functions as the algorithm can benefit by capturing the correlations between these bands.

\section{Plug-and-play ADMM}
\label{sec:pnp_admm}

The plug-and-play alternating direction method of multipliers (PnP ADMM) has emerged as a powerful framework for solving inverse problems by combining iterative optimization techniques with deep-learning-based priors. Originally developed for convex optimization problems with linear equality constraints \citep{Boyd2010}, ADMM decomposes the minimization process into sequential sub-problems, typically involving a data fidelity term and a regularization term, followed by an update of the dual variable. Previous works \citep{Venkatakrishnan2013, Sreehari2016, Chan2016} have interpreted these sub-steps as an inversion step followed by a denoising step, coupled via the augmented Lagrangian term and the dual variable. The PnP ADMM  approach extends this idea by replacing the proximal operator related to the prior with a deep neural network (DNN) trained as a denoiser \citep{Meinhardt2017, Bigdeli2017, Gupta2018, sureau2020}, allowing for greater flexibility in handling complex image priors. Compared to direct deep-learning-based inverse models, PnP ADMM offers several advantages: (1) it decouples the inversion step from the DNN, enabling the inclusion of additional convex constraints that can be efficiently handled via optimization, (2) it reduces the cost of learning by focusing solely on training a denoiser rather than multiple networks, as seen in unfolding approaches, and (3) by iterating between denoising and inversion, it ensures that the network output remains consistent with the observed data. In this work, we integrate the PnP ADMM framework from \cite{sureau2020} with the DRUNet denoiser from \cite{drunet}. Specifically, we employ DRUNet in the proximal update step of the PnP framework to enhance the denoising performance. 

\begin{algorithm*}
  \caption{Plug-and-Play ADMM algorithm to deconvolve a galaxy image, inspired by \cite{sureau2020}\label{Algo:ADMMPnP}}
  \begin{algorithmic}[1]
  \STATE \textbf{Initialize}: Set $\rho_0=1, \rho_{max}=10, \eta=0.5, \gamma=1.4, \Delta_0=0$, $\mathbf{x}^{(0)}=\mathbf{y}, \mathbf{z}^{(0)}=\mathbf{x}^{(0)}, \boldsymbol{\mu}^{(0)}=0, \epsilon$ 
  \FOR[\textbf{Main Loop}]{$k=0$ to $N_{iterations}$}
  \STATE \textbf{Deconvolution sub-problem}: $\mathbf{x}^{(k+1)}=FISTA(\textbf{y},\mathbf{x}^{(k)},\mathbf{z}^{(k)},\boldsymbol{\mu}^{(k)},\rho_k) \hfill \text{\citep{Beck2009}}$   
  \STATE \textbf{Denoising sub-problem}: $\mathbf{z}^{(k+1)} = N_{\theta} \left(\mathbf{x}^{(k+1)}+\boldsymbol{\mu}^{(k)}\right) \hfill (N_{\theta} = \text{DRUNet denoiser})$ 
  \STATE \textbf{Lagrange multiplier update}: $\boldsymbol{\mu}^{(k+1)}=\boldsymbol{\mu}^{(k)}+\left(\mathbf{x}^{(k+1)}-\mathbf{z}^{(k+1)}\right)$
  \STATE $\Delta_{k+1}=\frac{1}{\sqrt{n}}\left(||\mathbf{x}^{(k+1)}-\mathbf{x}^{(k)}||_2+||\mathbf{z}^{(k+1)}-\mathbf{z}^{(k)}||_2+||\boldsymbol{\mu}^{(k+1)}-\boldsymbol{\mu}^{(k)}||_2\right)$
  \IF{$\Delta_{k+1}\geq \eta \Delta_k$ \AND  $\rho_{k+1}\leq \rho_{max}$} 
      \STATE  $\rho_{k+1}=\gamma\rho_k$
  \ELSE
      \STATE  $\rho_{k+1}=\rho_k$
  \ENDIF

\ENDFOR
\RETURN $\left\{\mathbf{x}^{(k+1)}\right\}$
  \end{algorithmic}
\end{algorithm*}

\subsection{The proposed solution}
\label{pnp_admm_sol}

Considering the forward model described in Sect. \ref{forward_model1}, we defined the following loss functions, akin to Eqs. \ref{vloss}-\ref{zloss}, but incorporating an additional augmented Lagrangian term with the dual variable. These functions were then minimized using the algorithm outlined in Appendix \ref{pnp_admm_algo}.

\begin{align} 
    L_r(\mathbf{x}_r) &= \frac{1}{2} \left\Vert \frac{\mathbf{h}_r \ast \mathbf{x}_r - \mathbf{y}_r}{\sigma_r} \right\Vert_F^2 + \frac{\rho}{2} \left\Vert \mathbf{x}_r - \mathbf{z}_r + \mathbf{\mu}_r \right\Vert_F^2 + \lambda_{r} \textbf{K} \label{vloss_pnp}\\[10pt]
    L_i(\mathbf{x}_i) &= \frac{1}{2} \left\Vert \frac{\mathbf{h}_i \ast \mathbf{x}_i - \mathbf{y}_i}{\sigma_i} \right\Vert_F^2 + \frac{\rho}{2} \left\Vert \mathbf{x}_i - \mathbf{z}_i + \mathbf{\mu}_i \right\Vert_F^2 + \lambda_{i} \textbf{K} \label{iloss_pnp}\\[10pt]
    L_z(\mathbf{x}_z) &= \frac{1}{2} \left\Vert \frac{\mathbf{h}_z \ast \mathbf{x}_z - \mathbf{y}_z}{\sigma_z} \right\Vert_F^2 + \frac{\rho}{2} \left\Vert \mathbf{x}_z - \mathbf{z}_z + \mathbf{\mu}_z \right\Vert_F^2 + \lambda_{z} \textbf{K} \label{zloss_pnp}
\end{align}

\begin{align}
\text{where } \textbf{K} = \left\Vert \frac{\hspace{2pt} \mathbf{h}_{euc} \ast \sum\limits_{c\in\{r,i,z\}} \alpha_c \mathbf{x}_c - \mathbf{y}_{euc} }{\sigma_{euc}} \right\Vert_F^2. \label{eq:constr_pnp}
\end{align}

\noindent As before, the first terms in Eqs. \ref{vloss_pnp}-\ref{zloss_pnp} represent the data fidelity terms for each respective band, with $\sigma_r$, $\sigma_i$, and $\sigma_z$ being the corresponding noise maps. The second terms represent the augmented Lagrangian, incorporating the dual variable $\mathbf{z}$ to split the problem into two sub-problems: an inversion/deconvolution step followed by a denoising step. The third terms, as explained in Sect. \ref{forward_model}, are the constraining terms that enforce the condition that the sum of the Rubin $r$-, $i$-, and $z$-band images equals the Euclid VIS-band images. For the ADMM update, the parameter $\rho$ was manually tuned based on the approach from \cite{sureau2020} to strike a balance between quickly stabilizing the algorithm (with a higher $\rho$) and prioritizing the minimization of the data fidelity term in the early iterations (with a lower $\rho$). The values of all other hyperparameters have been previously described in Sects. \ref{forward_model} and \ref{hyperparam}.

\subsection{The algorithm}
\label{pnp_admm_algo}

The PnP ADMM algorithm inspired by \cite{sureau2020} is summarized in Table \ref{Algo:ADMMPnP}. The first step consists of solving the loss functions \ref{vloss_pnp}-\ref{zloss_pnp} alternatively at iteration $k$ using the accelerated iterative convex algorithm FISTA \citep{Beck2009}. In the second step, the DRUNet denoiser functions as a projector in the proximal update step, as described earlier. The final step regulates the augmented Lagrangian parameter, ensuring its increase when the optimization parameters exhibit insufficient change.

\subsection{Results}
\label{pnp_admm_results}

\begin{figure*}[h!]
\centering
    \subfigure[\makebox{}]
    {\label{subfig:mcdec1_pnp}\includegraphics[width=0.916\linewidth]{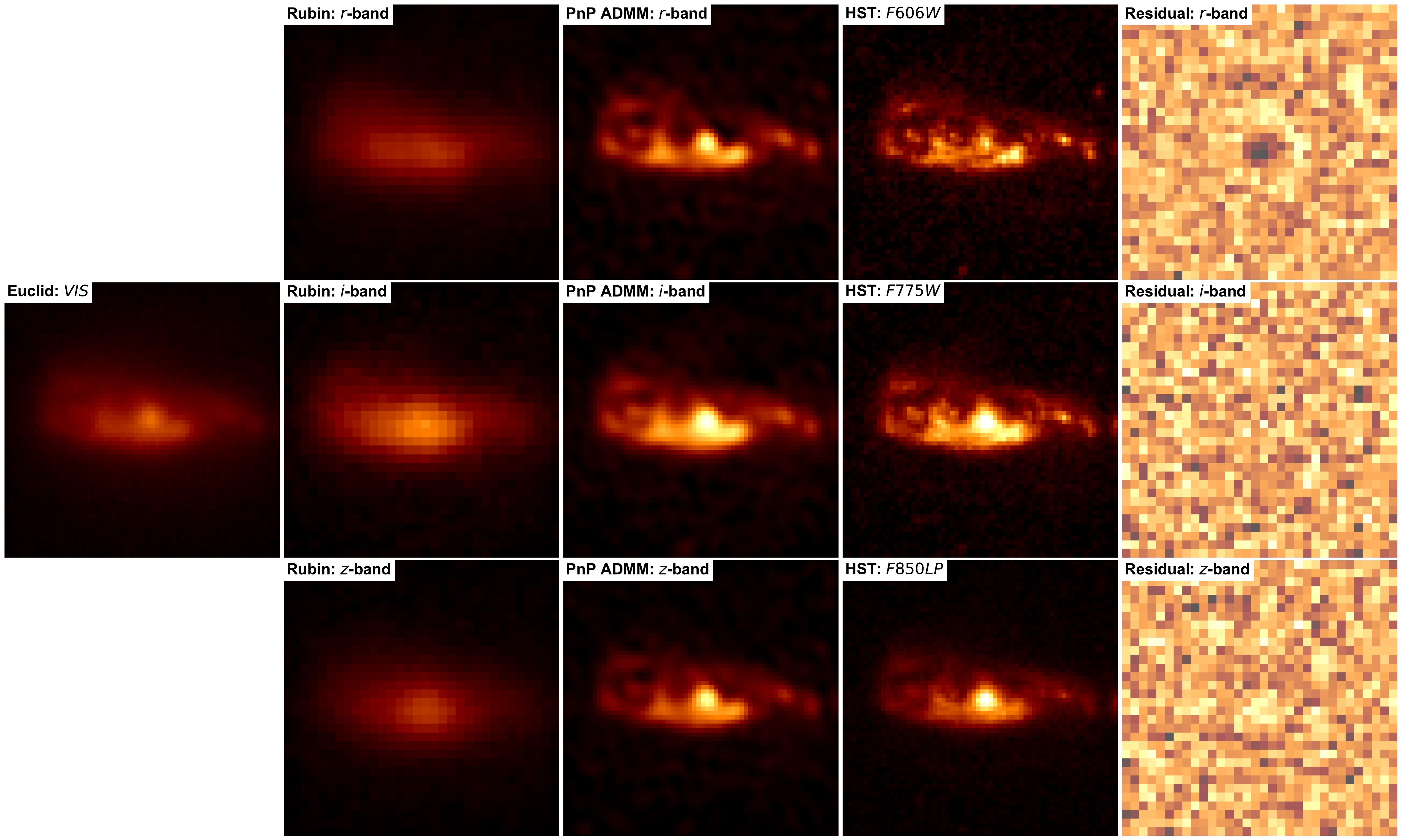}}\\
    
    \subfigure[\makebox{}]
    {\label{subfig:mcdec2_pnp}\includegraphics[width=0.916\linewidth]{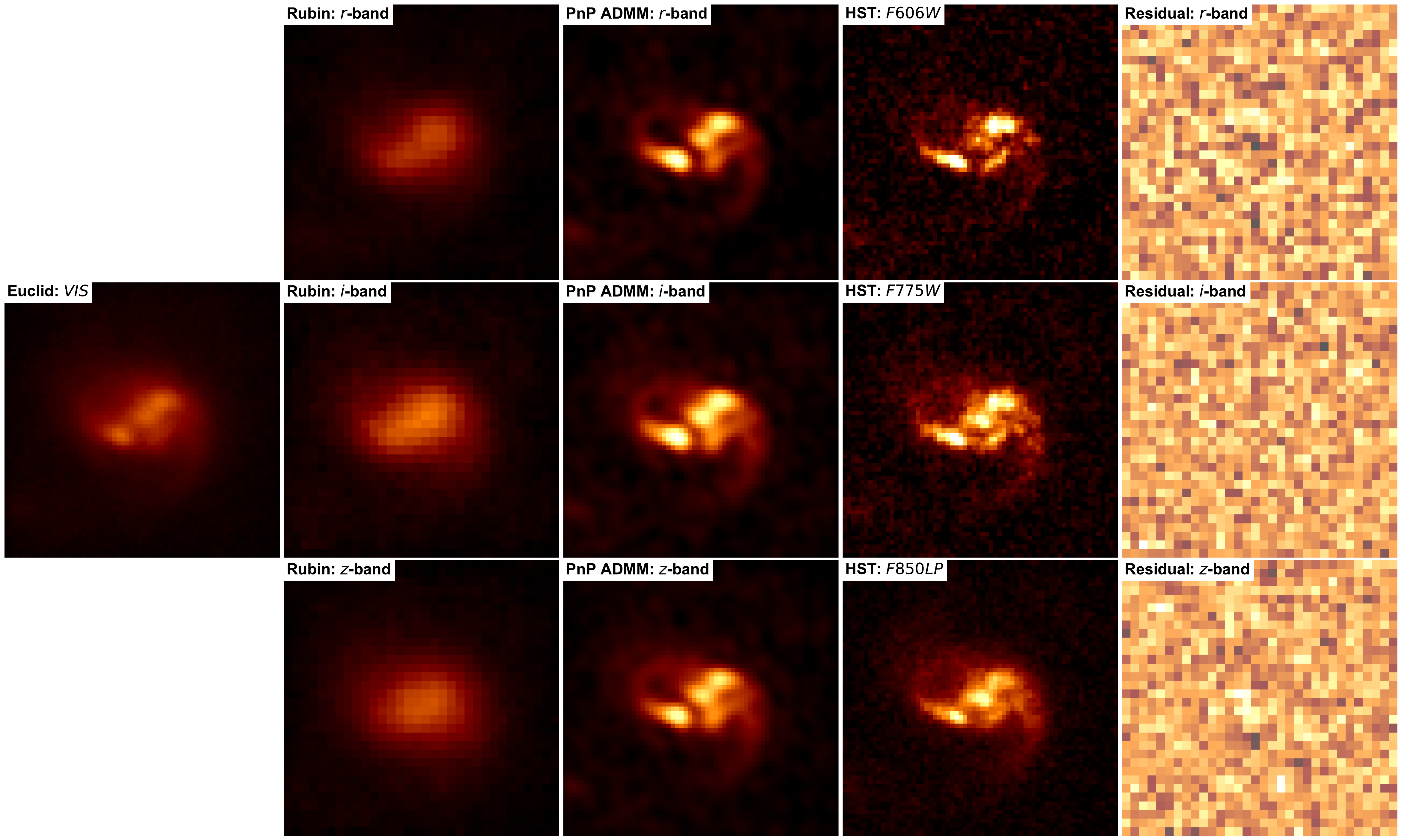}}
\caption{\label{fig:deconv_outputs_pnp} {Outputs of the PnP ADMM algorithm for the two galaxies shown in Fig. \ref{fig:deconv_outputs}, closely matching the results presented in the same figure. First column: Euclid VIS image.\ Second: Rubin simulations in the $r$, $i$, and $z$ bands. Third: Deconvolved outputs for the three bands. Fourth: Corresponding ground-truth HST images.\ Fifth: Residuals.}}
\end{figure*}

The algorithm simultaneously processed the noisy simulations from the three Rubin bands and the Euclid VIS band, along with their respective PSFs. These noisy images served as initializations or first guesses. The algorithm was run for 200 iterations, with convergence typically observed within $150$-$200$ iterations for all images in our dataset. Figure \ref{fig:loss_pnp} shows the convergence plot of the loss function for the deconvolved output in Fig. \ref{subfig:mcdec1_pnp}.

\begin{figure}[h!]
    \centering
    \label{fig:loss_pnp}\includegraphics[width=\linewidth]{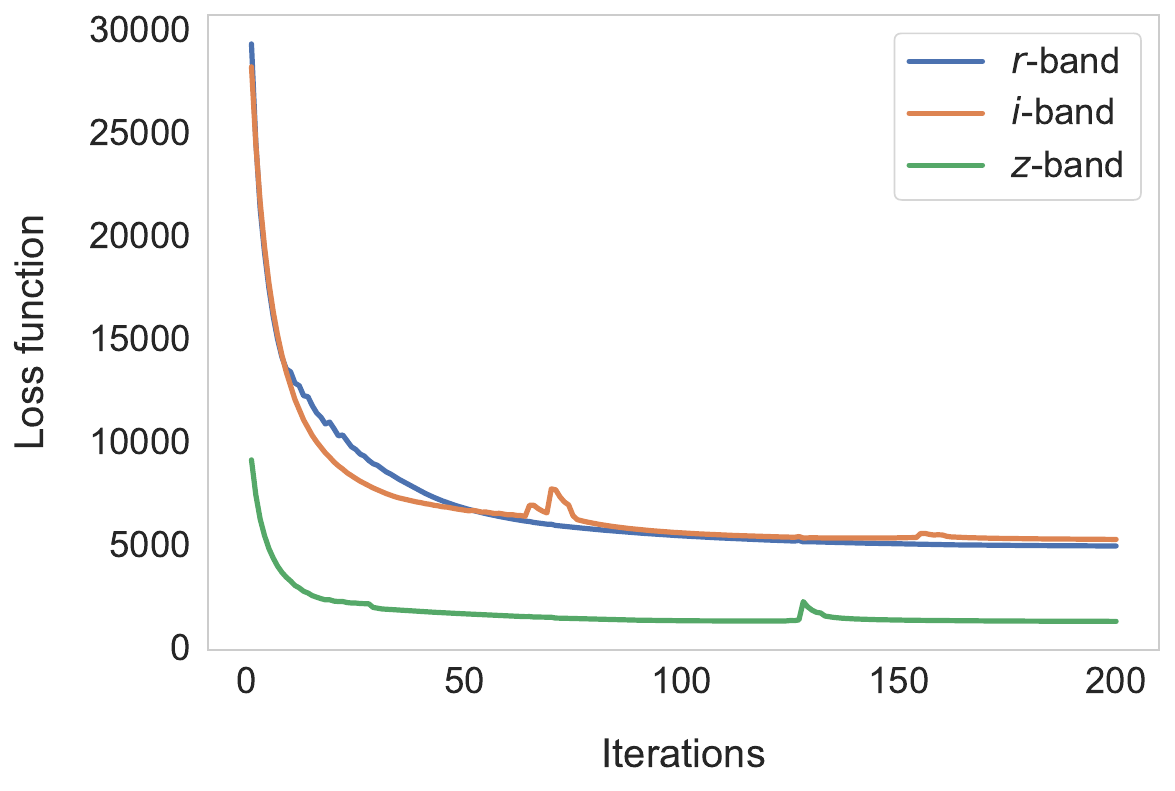}
    \caption{Loss function for the galaxy shown in Fig. \ref{subfig:mcdec1_pnp}. Convergence is achieved at around 200 iterations when the relative change in loss value is $<10^{-3}$ and the curve is flat.}
    \label{fig:loss_pnp}
\end{figure}

Figure \ref{fig:deconv_outputs_pnp} presents the same two examples of deconvolved images as shown in Fig. \ref{fig:deconv_outputs}. Qualitatively, the outputs for the two methods closely resemble each other. Compared to the original algorithm, there is a slight reduction in NMSE by approximately $0.26\%$. However, due to the additional computational steps involved in the iterative process and the proximal denoising step, PnP ADMM takes approximately 50 times longer to run than the original algorithm. This is consistent with the findings of \cite{sureau2020}. Hence, this significant increase in computational time makes it less practical for use on large datasets.

\end{appendix}
\end{document}